\documentclass[pra,aps,twocolumn,epsfig,nofootinbib,superscriptaddress]{revtex4}

\usepackage{amsmath}
\usepackage{amssymb}
\usepackage[dvips]{graphicx}
\usepackage{dcolumn}
\usepackage{bm}
\usepackage[driverfallback=dvips]{hyperref} 
\usepackage{setspace}
\usepackage{amsfonts}
\usepackage{rotating}
\usepackage{makeidx}
\usepackage{amsthm}
\usepackage{bbold}
\usepackage{float}
\usepackage{caption}
\usepackage{subfigure}

\begin{document}

\title{Fault-tolerant mixed boundary punctures on the toric code}

\author{Yao Shen}
\email[Corresponding author: ]{shenyaophysics@hotmail.com and shenyao@ppsuc.edu.cn}
\affiliation{School of Criminal Investigation, People's Public Security University
of China, Beijing 100038, China}

\author{Fu-Lin Zhang}
\affiliation{Department of Physics, School of Science, Tianjin University, Tianjin 300072, China}


\date{\today}

%

\begin{abstract}
Defects on the toric code,  a well-known exactly solvable Abelian anyon model, can exhibit non-Abelian statistical properties,
 which can be classified into punctures and twists.
Benhemou \emph{et al.} [Phys. Rev. A. 105, 042417 (2022)] introduced
a mixed boundary puncture model that integrates the advantages of both punctures and twists.
They proposed that non-Abelian properties could be realized in the symmetric subspace  $\{\left|++\right\rangle ,\left|--\right\rangle \}$.
This work demonstrates that the nontrivial antisymmetric subspace $\{\left|+-\right\rangle ,\left|-+\right\rangle \}$ also supports non-Abelian statistics.
The mixed boundary puncture model is shown to be fault-tolerant in both subspaces, offering resistance to collective dephasing noise and collective rotation noise.
In addition, we propose and validate a quantum information masking scheme within the three-partite mixed boundary puncture model.
%
%
\end{abstract}
%
%

\keywords{Toric code, Kitaev anyon, Ising anyon}
	
 \maketitle
	
\section{Introduction}

In 1997, Kitaev \cite{k00} proposed a topological code,
known as the toric code or surface code, depending on the boundary conditions.
Typically defined on a boundaryless surface,
the toric code's low-energy excitations are Abelian anyons\cite{k0,k1,k2,k3}.
The braiding group representation of Abelian anyons is a one-dimensional irreducible representation;
however, topological quantum computation (TQC) generally requires two-dimensional or higher representations to implement quantum gate operations.
This limitation makes the toric code, in its basic form, generally unsuitable for TQC.

%


%

Nevertheless, defects on the toric code can introduce non-Abelian statistics \cite{de,de1},
overcoming this limitation and making the toric code an important platform for  TQC \cite{key-2,key-3,key-4,m0,m1,m2,m3}.
The toric code includes both local (point-like) defects \cite{p1,p2,p3,p4,p5} and nonlocal (line-like or twist) defects \cite{t1,t2,t3,t4}.
Punctures are  local holes in the lattice,  whereas twists correspond to the endpoints of nonlocal domain walls.
Their behavior closely resembles that of Majorana zero modes \cite{u,u1,u2,u3,u4}.
Recently, Benhemou \emph{et al.} \cite{de1} studied the hybridization of these two defects,
effectively combining their advantages and demonstrating the presence of non-Abelian statistics in Ising anyons.
In this framework, information is encoded in nonlocal degrees of freedom, and universal quantum gates can be realized through braiding and fusion operations.


The toric code utilizes its topological properties to protect qubits from certain types of errors, making it a promising platform for fault-tolerant quantum computation.
First, toric code punctures exhibit a high fault-tolerance threshold of around $1\%$, meaning that increasing the code distance (i.e., enlarging the size of the surface code) can significantly reduce logical error rates \cite{key-21,key-22,key-23}.
Second, the toric code only requires parity checks between nearest neighbors, simplifying its implementation in physical systems \cite{key-2,key-3,key-4}.
As a result, topological quantum systems are highly competitive due to their combination of a high fault-tolerance threshold, accuracy, and minimal qubit overhead.
The toric code plays a crucial role in TQC, offering unique encoding schemes and robust error correction.
Therefore, it is not only a key component in TQC but also an important focus of research in quantum computing.


This work builds upon the research of Benhemou \emph{et al.} \cite{de1} by demonstrating mixed boundary punctures and realizing non-Abelian statistics in the nontrivial antisymmetric subspace $\{\left|+-\right\rangle ,\left|-+\right\rangle \}$,
while Benhemou \emph{et al.} proposed the realization in the symmetric subspace $\{\left|++\right\rangle,\left|--\right\rangle \}$.
These two subspaces are independent of each other, distinguished by their respective realization methods within the lattice model.
 We show that both systems are immune to two typical types of noise -- collective dephasing and collective rotation -- within their respective subspaces. Consequently, these two subspaces of the mixed boundary punctures on the toric code are fault-tolerant.
 This conclusion highlights the significant potential of non-Abelian statistics in the context of the Abelian lattice model for TQC.
 Additionally, we propose and validate a quantum information masking scheme applicable to the three-partite mixed boundary puncture model.


In the second section, we present the Abelian and non-Abelian anyon models, along with the mixed boundary punctures model.
Section III discusses the robustness of the mixed boundary punctures model against two types of collective noise.
In Sec. IV, we  design  the information masking scheme within the mixed boundary punctures model.
Finally, we provide a summary in Sec. V.


\section{Mixed boundary punctures}

\subsection{Anyon models}

The unitary modular tensor category is the algebraic structure describing
the general anyon system. The total anyonic space consists of all Hilbert
subsystem spaces and the fusion spaces of all charges\cite{k00,k0,k1,k2,k3}.
\begin{equation}
H_{ab}^{c}=\underset{ab}{\oplus}H_{a}\otimes H_{b}\otimes V_{ab}^{c},
\end{equation}
where $H_{a}$ represents the Hilbert subspace of charge $a$, and 
$V_{ab}^{c}$ is the fusion space containing the fusion rules. In
the fusion space $V_{ab}^{c}$ , the fusion rule seems like a superposition.
\begin{equation}
a\otimes b=\underset{c\in L}{\oplus}N_{ab}^{c}c,
\end{equation}
where $a,b\in L$ and $N_{ab}^{c}$ is the possible ways of fusing
$a$ and $b$. The matrix relating two different bases of the splitting
trees is F-symbol. The braiding action is called R-symbol (see Fig 1.).
The algebraic model of anyons is composed of the fusion rules, F-symbols,
R-symbols, and some other data of the unitary modular tensor category.
In this paper, non-Abelian statistics of Ising anyon is realized using
the mixed boundary punctures on the toric code of an Abelian anyon
model.

\begin{figure}
\includegraphics[scale=0.65]{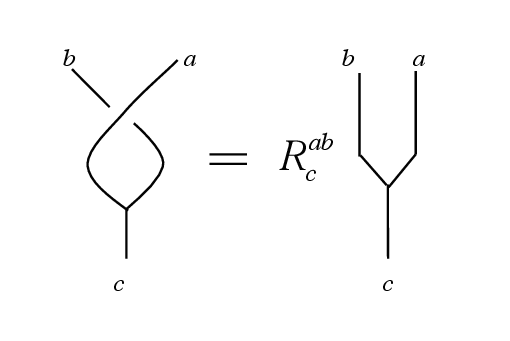}\caption{The braiding operation $R_{c}^{ab}$.}
\end{figure}

Wilczek first discovered the braiding group of a kind of quasiparticle,
which gave an additional Aharonov-Bohm phase after the braiding operation\cite{key-29,key-30}.
These quasiparticles are called Abelian anyons. Immediately after,
Kitaev introduced a marvelous exactly solvable model \cite{k00,k0,k1,k2}.
The Kitaev toric code can be viewed as a simplification and generalization
of a quantum $Z_{2}$ gauge theory. It is a square lattice grid with
each edge or vertex hosting a spin-1/2 particle. The grids satisfie
periodic boundary conditions forming a torus-like structure. The excitations
are Abelian anyons. The vacuum $1$, electric charge $e$, magnetic
flux $m$ and fermion $\varepsilon=e\times m$ are four superselection
sectors of the Abelian anyons (the case of two mutual antiparticles
is ignored here). For Abelian anyons, Chern number $c=0,8$ and $c=\pm4$
are two categories which mod $16$. The parameters of the simplest
Abelian anyons are those with Chern number $c=0$, topology spin $\theta=1$
and Frobenius-Schur indicator $\kappa=1$. The Abelian anyons are
mod 2, and the fusion rules read
\begin{equation}
\begin{array}{c}
e\times e=m\times m=\varepsilon\times\varepsilon=1,\\
\varepsilon\times e=m,\:\varepsilon\times m=e,\:e\times m=\varepsilon.
\end{array}\label{eq:1}
\end{equation}
According to the definition of braiding operation (see Fig 1.), the braiding (or
exchange) rules of those four superselection sectors are
\begin{equation}
\begin{array}{c}
R_{\varepsilon}^{em}=-R_{\varepsilon}^{me}=1,\\
R_{e}^{\varepsilon m}=-R_{e}^{m\varepsilon}=1,\\
R_{m}^{e\varepsilon}=-R_{m}^{e\varepsilon}=1,\\
R_{1}^{ee}=R_{1}^{mm}=-R_{1}^{\varepsilon\varepsilon}=1.
\end{array}\label{eq:2}
\end{equation}
It is worth noting that the results given by different orders of superscripts
on operators are also different. For example, $R_{\varepsilon}^{em}=- R_{\varepsilon}^{me}=1$.
This is not trivial, the representation of another intermediate statistical-Gentile
statistics can give a much more clearer physical image for the braiding.
In the Gentile statistics representation space, $R_{\varepsilon}^{em}$
and $R_{\varepsilon}^{me}$ are mapped into two mutual conjugate spaces
\cite{me}. This kind of anyons are called the Abelian $1/2$-anyons,
because the statistical parameter is $1/2$ for winding number $1$.

Ising anyons are the simplest non-Abelian anyons. Their Chern number
are $c=1$. Similarly, there are three categories of excitations:
the vacuum $1$, the Majorana fermion $\psi$ and the vortex $\sigma$.
For non-Abelian anyons, the topological spin and the Frobenius-Schur
indicators of vortices are divided into $8$ pieces $\theta_{\sigma}=exp(i\pi c/8)$
and $\varkappa_{\sigma}=(-1)^{(c^{2}-1)/8}$ , besides $\theta_{1}=1,\:\theta_{\psi}=-1,\:\varkappa_{1}=\varkappa_{\psi}=1$.
For Ising anyon, $\theta_{\sigma}=exp(i\pi/8)$ and $\varkappa_{\sigma}=\varkappa=1$.
The fusion rules of non-Abelian Ising anyons are
\begin{equation}
\psi\times\psi=1,\quad\psi\times\sigma=\sigma,\quad\sigma\times\sigma=1+\psi.\label{eq:3}
\end{equation}
And the braiding rules of Ising anyons give
\begin{equation}
\begin{array}{cc}
R_{1}^{\psi\psi}=-1,\: & R_{1}^{\sigma\sigma}=\varkappa e^{-\frac{i\pi c}{8}}=e^{-\frac{i\pi}{8}},\\
R_{\sigma}^{\psi\sigma}=R_{\sigma}^{\sigma\psi}=-i^{c}=-i, & \:R_{\psi}^{\sigma\sigma}=\varkappa e^{\frac{i3\pi c}{8}}=e^{\frac{i3\pi}{8}}.
\end{array}\label{eq:4}
\end{equation}

\subsection{Defects on toric code}

The toric code can be defined on lattice with one qubit at each vertex.
The Hamiltonian can be expressed as
\begin{equation}
H=\underset{f}{\sum}A_{f}+\underset{f}{\sum}B_{f},
\end{equation}
with stabilizers
\begin{equation}
A_{f}=\underset{j\in\partial f}{\prod}X_{j},\;B_{f}=\underset{j\in\partial f}{\prod}Z_{j},
\end{equation}
where $\partial f$ is the qubits that connected to the face \cite{de1}. The
information can be encoded on the toric code by introducing defects
on the model surface. Two common kinds of defects are punctures and twists.
Measurements of the stabilizers could disentangle the spin system and
create the punctures on the lattice. The logical qubits can be encoded
using the parity of the puncture anyons. When Pauli $Z$ stabilizers
are measured, electric charge $e$ is excited (the boundary of the
puncture is solid). When Pauli $X$ stabilizers are measured, magnetic
flux $m$ is excited (the boundary of the puncture is dash) (see Fig.2). Twists
are extrinsic defects on toric code. They can be represented by five
stabilizer operators $XZYXZ$ which cases a dislocation at the endpoints
of the lattice. The anyon braiding around a twist could exchange the
type of $e$ and $m$, while leave $\psi$, braiding and fusion rules
invariant \cite{de,de1}. 

In 2022, A. Benhemou et al. designed a new concept of mixed boundary
punctures which combined the punctures and twists (Fig.2)\cite{de1}.
In Fig.2, half solid and half dashed boundaries are juxtaposed and
connected with a pair of twists represented by two crosses. In Ref.[\cite{de}], B. J. Brown had proven the equivalence between the corners of the toric code and twist defects.
This mixed
boundary corresponds to the measurements of both $X$ and $Z$ stabilizers.
The twist applies a Pauli $Y$ operator at each intersection qubit,
then the mixed boundary punctures are completed. The strings connected
two mixed boundary puncture indicate that two $e$ or $m$ can be condensed
in those two punctures. The string which connects two solid sides
represents two $e$ are condensed, while two $m$ correspond to the
string connecting two dashed sides (Fig.3)\cite{de1}.

\begin{figure}
\includegraphics[scale=0.3]{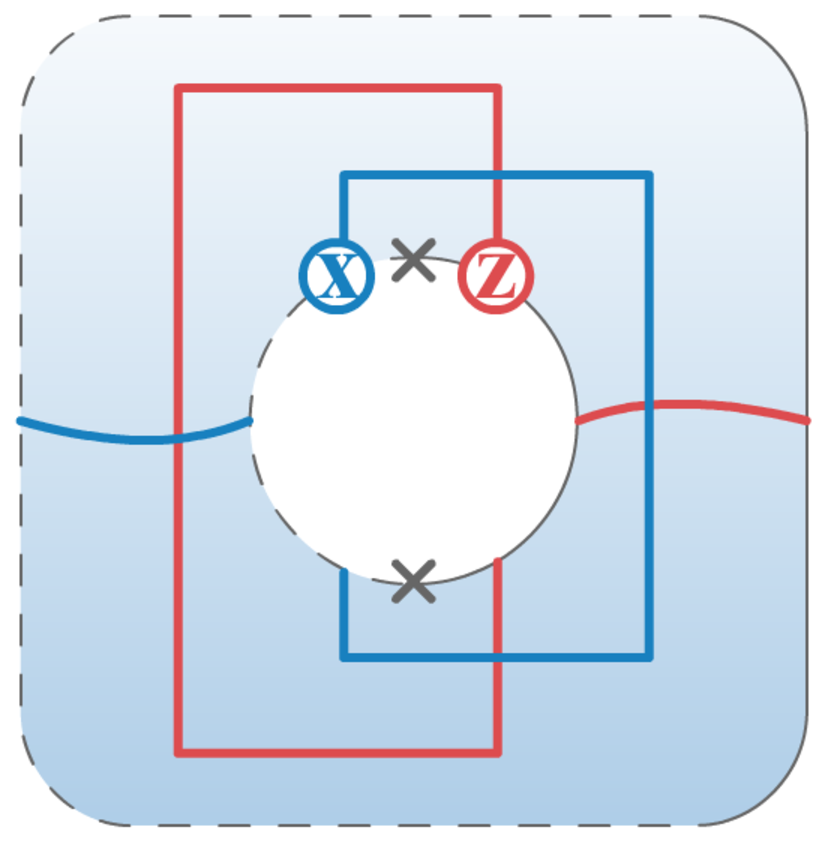}\caption{The mixed boundary punctures model. $Z$ type strings (red) and $X$ type strings (blue) terminate at solid and dashed boundaries respectively\cite{de1}.
Pauli $Z$ and $X$ operators (loops) stabilize this defect. The two horizontal curves represent the stabilizers connecting the boundary and the puncture.The two black crosses at the intersection of the solid boundary and the dashed boundary represent two twists.}
\end{figure}

\begin{figure}
\includegraphics[scale=0.3]{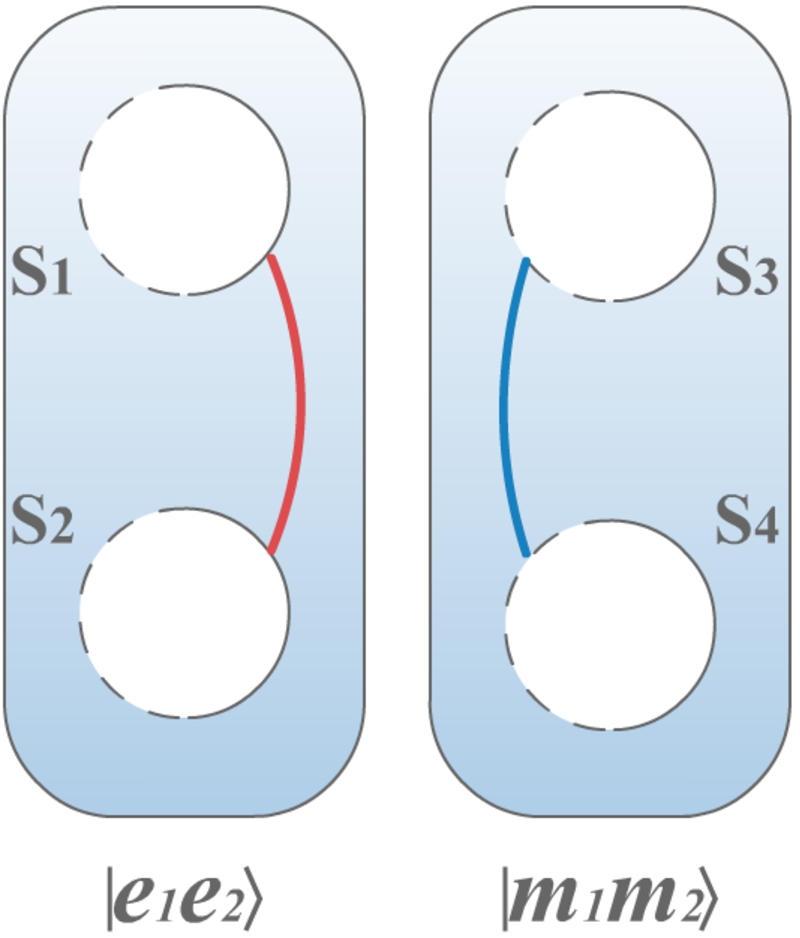}\caption{The states of the mixed boundary punctures. Red string is $Z$ type and blue string is $X$ type. The strings connect the matching boundaries of the puncture pairs.The $e$ and $m$ particles are condensed inside the punctures which means the states of the system are still ground states.}
\end{figure}

\subsection{Subspaces of mixed boundary punctures}


As A. Benhemou et al. mentioned\cite{de1}, the logical qubits are
the superposition states in Fig.3

\begin{equation}
\left|s_{1}s_{2},\pm\right\rangle =\frac{\left|e_{1}e_{2}\right\rangle \pm\left|m_{1}m_{2}\right\rangle }{\sqrt{2}}.\label{eq:0}
\end{equation}
They constructed symmetric states to satisfy the the fusion matrix
$F$ and braiding evolution matrix $B$ of Ising anyons
\begin{equation}
F=\frac{1}{\sqrt{2}}\left(\begin{array}{cc}
1 & 1\\
1 & -1
\end{array}\right),\label{eq:a}
\end{equation}
and
\begin{equation}
B=FR^{2}F^{-1}=e^{-i\frac{\pi}{4}}\left(\begin{array}{cc}
0 & 1\\
1 & 0
\end{array}\right).\label{eq:b}
\end{equation}

Actually, we discover that the antisymmetric construction also satisfies
the Ising statistics.


\emph{Symmetric subspace.--}
In Benhemou's research they only gave one symmetric construction,
they thought the antisymmetric case might not suitable for Ising statistics\cite{de1}.
The basic units of the Ising statistics on toric code are the joint
states of two symmetric states $\left|(s_{1}s_{2},+)(s_{1}s_{2},+)\right\rangle $
or $\left|(s_{1}s_{2},-)(s_{1}s_{2},-)\right\rangle $. They proved
that this symmetric construction satisfied the fusion matrix $F$
and braiding evolution matrix $B$ of Ising anyons. In Fig.4, braiding
$s_{1}$ around $s_{3}$ brings different results. When $s_{1}$ and
$s_{3}$ condense the same kind of quasiparticles, braiding create
the same operator loops enclosing the punctures to the strings. For
example, in Fig.4 (a) and (b), $s_{1}$ and $s_{3}$ are both $e$($m$)
particles, the strings are $Z$($X$)-type stabilizers. Braiding $s_{1}$
around $s_{3}$ gives both $s_{1}$ and $s_{3}$ a $Z$($X$) loop
crossing the $Z$($X$) string. When $s_{1}$ and $s_{3}$ condense
different kind of quasiparticles, braiding creates different operator
loops enclosing the punctures to the strings. For instance, in Fig.4
(c) and (d), $s_{1}$ and $s_{3}$ are $e$($m$) and $m(e)$ respectively,
the strings are $Z$($X$)-type and $X(Z)$-type stabilizers. Braiding
$s_{1}$ around $s_{3}$ gives $s_{1}$ and $s_{3}$ $X(Z)$ and $Z$($X$)
loops crossing the $Z$($X$) and $X(Z)$ strings.

Pauli $X$ operation corresponds to two braiding operations $B^{2}$,
The realization of Pauli $Z$ operation is creating a pair of $\psi$
in $s_{3}$ and $s_{4}$ and transmitting one particle of the pair
to $s_{1}$ and $s_{2}$ respectively (Fig.5(a))\cite{de1}.

\begin{figure}
\includegraphics[scale=0.45]{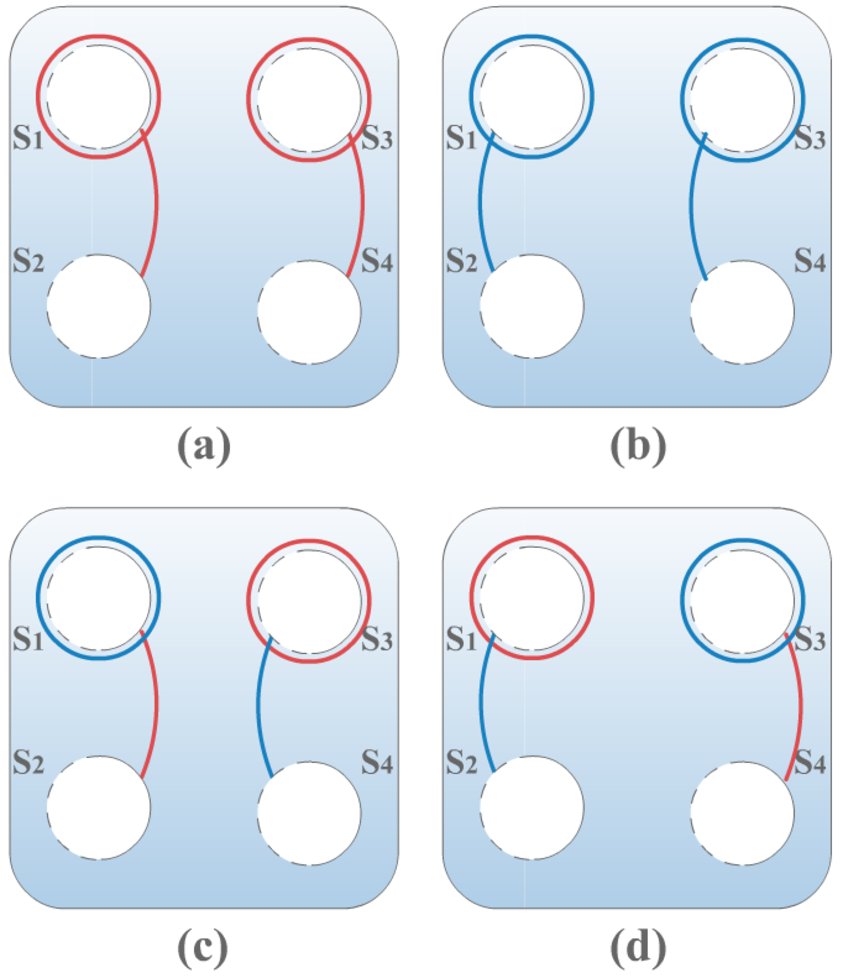}\caption{The basic units of the Ising statistics on toric code. Braiding $s_{1}$ around $s_{3}$ brings different results. Panels (a) and (b) is trivial, loops and strings are same type. In panels (c) and (d), loops and strings are different types, which give minus signs.}
\end{figure}

\begin{figure}
\includegraphics[scale=0.45]{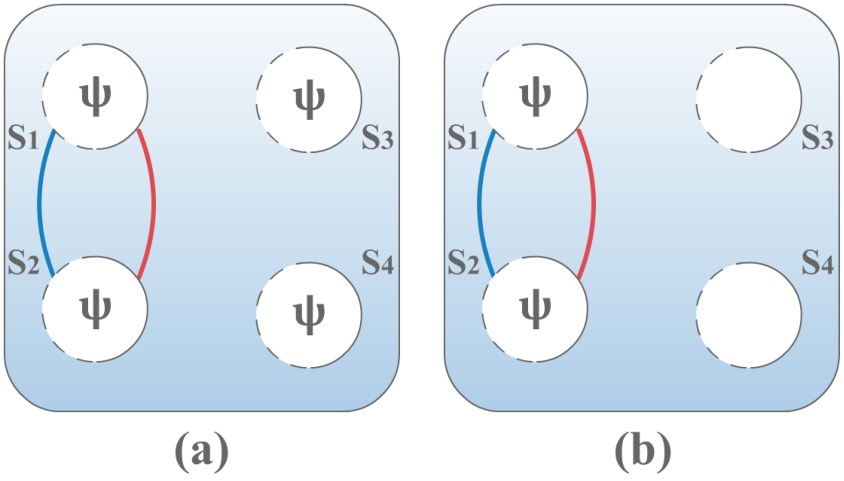}\caption{The realization of Pauli $Z$ operation. Panels (a) is the Pauli $Z$ in the symmetric subspace, while panels (b) is the Pauli $Z$ in the antisymmetric subspace.}
\end{figure}


\emph{Antisymmetric subspace.--}
In this part, we discuss the antisymmetric construction of the mixed
boundary punctures on toric code. We prove that this mixed boundary
punctures model also satisfy the Ising statistics in the antisymmetric
subspace $\{\left|(s_{1}s_{2},+)(s_{1}s_{2},-)\right\rangle ,\left|(s_{1}s_{2},-)(s_{1}s_{2},+)\right\rangle \}$.
As mentioned earlier, we need at least four mixed boundary punctures
to show the Ising statistics. In the antisymmetric subspace, the basic
vectors are
\begin{equation}
\begin{array}{c}
\left|(s_{1}s_{2},+)(s_{1}s_{2},-)\right\rangle \equiv\left|+-\right\rangle \\
=\frac{1}{2}(\left|e_{1}e_{2}e_{3}e_{4}\right\rangle -\left|e_{1}e_{2}m_{3}m_{4}\right\rangle \\
+\left|m_{1}m_{2}e_{3}e_{4}\right\rangle -\left|m_{1}m_{2}m_{3}m_{4}\right\rangle ),
\end{array}
\end{equation}
\begin{equation}
\begin{array}{c}
\left|(s_{1}s_{2},-)(s_{1}s_{2},+)\right\rangle \equiv\left|-+\right\rangle \\
=\frac{1}{2}(\left|e_{1}e_{2}e_{3}e_{4}\right\rangle +\left|e_{1}e_{2}m_{3}m_{4}\right\rangle \\
-\left|m_{1}m_{2}e_{3}e_{4}\right\rangle -\left|m_{1}m_{2}m_{3}m_{4}\right\rangle ).
\end{array}
\end{equation}
It can be easily proved that the action of the fusion matrix is
\begin{equation}
F\left|+-\right\rangle =\frac{1}{\sqrt{2}}(\left|e_{1}e_{2}e_{3}e_{4}\right\rangle -\left|m_{1}m_{2}m_{3}m_{4}\right\rangle )=\left|1_{13}1_{24}\right\rangle ,
\end{equation}
\begin{equation}
F\left|-+\right\rangle =\frac{1}{\sqrt{2}}(-\left|e_{1}e_{2}m_{3}m_{4}\right\rangle +\left|m_{1}m_{2}e_{3}e_{4}\right\rangle )=\left|\psi_{13}\psi_{24}\right\rangle .
\end{equation}
Where we define the vacuum and the Majorana fermion states
\begin{equation}
\begin{array}{c}
\left|1_{13}1_{24}\right\rangle =\frac{1}{\sqrt{2}}(\left|e_{1}e_{2}e_{3}e_{4}\right\rangle -\left|m_{1}m_{2}m_{3}m_{4}\right\rangle ),\\
\left|\psi_{13}\psi_{24}\right\rangle =\frac{1}{\sqrt{2}}(-\left|e_{1}e_{2}m_{3}m_{4}\right\rangle +\left|m_{1}m_{2}e_{3}e_{4}\right\rangle ),
\end{array}
\end{equation}
so the basic vectors
\begin{equation}
\begin{array}{c}
\left|+-\right\rangle =\frac{1}{\sqrt{2}}(\left|1_{13}1_{24}\right\rangle +\left|\psi_{13}\psi_{24}\right\rangle ),\\
\left|-+\right\rangle =\frac{1}{\sqrt{2}}(\left|1_{13}1_{24}\right\rangle -\left|\psi_{13}\psi_{24}\right\rangle ).
\end{array}
\end{equation}

When the braiding operation acts on those two states, we have
\begin{equation}
B_{13}^{2}\left|+-\right\rangle =\left|-+\right\rangle ,\;B_{13}^{2}\left|-+\right\rangle =\left|+-\right\rangle ,
\end{equation}
where $B_{13}^{2}$ means braiding particle $1$ around particle $3$,
and it is a Pauli $X$ operation obviously. But the Pauli $Z$ operation
is different from the symmetric case. In the antisymmetric subspace,
Pauli $Z$ operation is the process of creating a pair of $\psi$
in $s_{1}$ and $s_{2}$ (Fig.5(b)). After that, the states are
\begin{equation}
Z\left|+-\right\rangle =\left|+-\right\rangle ,\;Z\left|-+\right\rangle =-\left|-+\right\rangle .
\end{equation}
What calls for special attention is that $\psi$ is the fermion $\varepsilon$
in toric code, their single particle fusion and braiding rule follow
Eq.(\ref{eq:1}), but the total effect of the states $\{\left|+-\right\rangle ,\left|-+\right\rangle \}$
satisfies the Ising statistics Eq.(\ref{eq:a}) and Eq.(\ref{eq:b}).

\section{Fault-tolerant properties}

The mixed boundary punctures model is a simulation of Ising anyon statistical
properties, not real Ising anyon. Non-Abelian anyons are intrinsic
fault-tolerant, because the Hilbert space of non-Abelian anyons is
composed of the local subspace and the logical subspace. Errors only
exist in the local subspace which is isolated to the logical subspace.
The logical subspace is determined by the conjugacy class of finite
group and the irreducible representation of its centralizer. Local
operations can not access to the logical subspace, quantum information
is protected from local errors. Whether this simulation is immune
to noise? We consider two typical noises: the collective dephasing
noise and the collective rotation noise. We confirm to the extent
that the symmetric subspace $\{\left|++\right\rangle ,\left|--\right\rangle \}$
and the antisymmetric subspace $\{\left|+-\right\rangle ,\left|-+\right\rangle \}$
of the mixed boundary punctures model can be regarded as two logical
subspaces of Ising anyons.

\subsection{Dephasing noise}

We assume $\left|0\right\rangle =\left|e\right\rangle $ and $\left|1\right\rangle =\left|m\right\rangle $
on toric code. The collective dephasing noise gives
\begin{equation}
U_{d}\left|e\right\rangle =\left|e\right\rangle ,\;U_{d}\left|m\right\rangle =e^{i\phi(t)}\left|m\right\rangle ,\label{eq:d1}
\end{equation}
where $\phi(t)$ is a random phase that varies over time.
\begin{itemize}
\item The symmetric subspace $\{\left|++\right\rangle ,\left|--\right\rangle \}$
\end{itemize}
According to Eq.(\ref{eq:0}) and Eq.(\ref{eq:d1}), in two qubits
case, we have
\begin{equation}
\underset{2}{\otimes}U_{d}\left|\pm\right\rangle =\frac{1}{\sqrt{2}}(\left|e_{1}e_{2}\right\rangle \pm e^{i2\phi}\left|m_{1}m_{2}\right\rangle )\equiv\left|\tilde{\pm}\right\rangle .\label{eq:d2}
\end{equation}
In the symmetric subspace, when the collective dephasing noise operator
acts on the basic unit of the Ising anyons ($4$ mixed boundary punctures),
the states become
\begin{equation}
\begin{array}{c}
\underset{4}{\otimes}U_{d}\left|\pm\pm\right\rangle =\frac{1}{2}(\left|e_{1}e_{2}e_{3}e_{4}\right\rangle \pm e^{i2\phi}\left|e_{1}e_{2}m_{3}m_{4}\right\rangle \\
\pm e^{i2\phi}\left|m_{1}m_{2}e_{3}e_{4}\right\rangle +e^{i4\phi}\left|m_{1}m_{2}m_{3}m_{4}\right\rangle \equiv\left|\tilde{\pm}\tilde{\pm}\right\rangle .
\end{array}
\end{equation}
By calculation, it can be proven that $\left\langle \tilde{+}\tilde{+}\right.\left|\tilde{-}\tilde{-}\right\rangle =0$
and $\left\langle \tilde{\pm}\tilde{\pm}\right.\left|\tilde{\pm}\tilde{\pm}\right\rangle =1$.
The system in the symmetric subspace is immune to the collective dephasing
noise.
\begin{itemize}
\item The antisymmetric subspace $\{\left|+-\right\rangle ,\left|-+\right\rangle \}$
\end{itemize}
Similar to the symmetric subspace, under the effect of the collective
dephasing noise, the four qubits states become
\begin{equation}
\begin{array}{c}
\underset{4}{\otimes}U_{d}\left|\pm\mp\right\rangle =\frac{1}{2}(\left|e_{1}e_{2}e_{3}e_{4}\right\rangle \mp e^{i2\phi}\left|e_{1}e_{2}m_{3}m_{4}\right\rangle \\
\pm e^{i2\phi}\left|m_{1}m_{2}e_{3}e_{4}\right\rangle -e^{i4\phi}\left|m_{1}m_{2}m_{3}m_{4}\right\rangle \equiv\left|\tilde{\pm}\tilde{\mp}\right\rangle .
\end{array}
\end{equation}
It is easy to verify that $\left\langle \tilde{+}\tilde{-}\right.\left|\tilde{-}\tilde{+}\right\rangle =0$
and $\left\langle \tilde{\pm}\tilde{\mp}\right.\left|\tilde{\pm}\tilde{\mp}\right\rangle =1$.
The system in the antisymmetric subspace is immune to the collective
dephasing noise.

\subsection{Rotation noise}

The collective rotation noise affects the states as
\begin{equation}
\begin{array}{c}
U_{r}\left|e\right\rangle =\cos\theta\left|e\right\rangle +\sin\theta\left|m\right\rangle ,\\
U_{r}\left|m\right\rangle =-\sin\theta\left|e\right\rangle +\cos\theta\left|m\right\rangle ,
\end{array}\label{eq:d3}
\end{equation}
where $\theta(t)$ is also a random phase that varies over time.
\begin{itemize}
\item The symmetric subspace $\{\left|++\right\rangle ,\left|--\right\rangle \}$
\end{itemize}
In the symmetric subspace, according to Eq.(\ref{eq:d3}), it's not
difficult to draw a conclusion that $\left|++\right\rangle $ is invariant
under the collective rotation noise

\begin{equation}
\begin{array}{c}
\underset{4}{\otimes}U_{r}\left|++\right\rangle =\frac{1}{2}(\left|e_{1}e_{2}e_{3}e_{4}\right\rangle +\left|e_{1}e_{2}m_{3}m_{4}\right\rangle \\
+\left|m_{1}m_{2}e_{3}e_{4}\right\rangle +\left|m_{1}m_{2}m_{3}m_{4}\right\rangle \equiv\left|\tilde{+}\tilde{+}\right\rangle .
\end{array}
\end{equation}

Another state is slightly more complex that
\begin{equation}
\begin{array}{c}
\underset{4}{\otimes}U_{r}\left|--\right\rangle =\frac{1}{2}[\prod \limits_{p=1}^{4}(\cos\theta\left|e_{p}\right\rangle +\sin\theta\left|m_{p}\right\rangle )\\
-\prod \limits_{p=1}^{2}\prod \limits_{q=3}^{4}(\cos\theta\left|e_{p}\right\rangle +\sin\theta\left|m_{p}\right\rangle )(-\sin\theta\left|e_{q}\right\rangle +\cos\theta\left|m_{q}\right\rangle )\\
-\prod \limits_{p=1}^{2}\prod \limits_{q=3}^{4}(-\sin\theta\left|e_{p}\right\rangle +\cos\theta\left|m_{p}\right\rangle )(\cos\theta\left|e_{q}\right\rangle +\sin\theta\left|m_{q}\right\rangle )\\
+\prod \limits_{p=1}^{4}(-\sin\theta\left|e_{p}\right\rangle +\cos\theta\left|m_{p}\right\rangle )]\equiv\left|\tilde{-}\tilde{-}\right\rangle .
\end{array}
\end{equation}

After a series of derivations, we have proved that $\left\langle \tilde{+}\tilde{+}\right.\left|\tilde{-}\tilde{-}\right\rangle =0$
and $\left\langle \tilde{\pm}\tilde{\pm}\right.\left|\tilde{\pm}\tilde{\pm}\right\rangle =1$
under the collective rotation noise. Therefore, symmetric subspaces
can resist the collective rotation noise.
\begin{itemize}
\item The antisymmetric subspace $\{\left|+-\right\rangle ,\left|-+\right\rangle \}$
\end{itemize}
The case in the symmetric subspace, under the actions of the collective
rotation noise, the basic vectors become
\begin{equation}
\begin{array}{c}
\underset{4}{\otimes}U_{r}\left|+-\right\rangle =\frac{1}{2}[\prod \limits_{p=1}^{4}(\cos\theta\left|e_{p}\right\rangle +\sin\theta\left|m_{p}\right\rangle )\\
-\prod \limits_{p=1}^{2}\prod \limits_{q=3}^{4}(\cos\theta\left|e_{p}\right\rangle +\sin\theta\left|m_{p}\right\rangle )(-\sin\theta\left|e_{q}\right\rangle +\cos\theta\left|m_{q}\right\rangle )\\
+\prod \limits_{p=1}^{2}\prod \limits_{q=3}^{4}(-\sin\theta\left|e_{p}\right\rangle +\cos\theta\left|m_{p}\right\rangle )(\cos\theta\left|e_{q}\right\rangle +\sin\theta\left|m_{q}\right\rangle )\\
-\prod \limits_{p=1}^{4}(-\sin\theta\left|e_{p}\right\rangle +\cos\theta\left|m_{p}\right\rangle )]\equiv\left|\tilde{+}\tilde{-}\right\rangle ,
\end{array}
\end{equation}
\begin{equation}
\begin{array}{c}
\underset{4}{\otimes}U_{r}\left|-+\right\rangle =\frac{1}{2}[\prod \limits_{p=1}^{4}(\cos\theta\left|e_{p}\right\rangle +\sin\theta\left|m_{p}\right\rangle )\\
+\prod \limits_{p=1}^{2}\prod \limits_{q=3}^{4}(\cos\theta\left|e_{p}\right\rangle +\sin\theta\left|m_{p}\right\rangle )(-\sin\theta\left|e_{q}\right\rangle +\cos\theta\left|m_{q}\right\rangle )\\
-\prod \limits_{p=1}^{2}\prod \limits_{q=3}^{4}(-\sin\theta\left|e_{p}\right\rangle +\cos\theta\left|m_{p}\right\rangle )(\cos\theta\left|e_{q}\right\rangle +\sin\theta\left|m_{q}\right\rangle )\\
-\prod \limits_{p=1}^{4}(-\sin\theta\left|e_{p}\right\rangle +\cos\theta\left|m_{p}\right\rangle )]\equiv\left|\tilde{-}\tilde{+}\right\rangle .
\end{array}
\end{equation}
We can still prove that the inner product remains invariant under
the collective rotation noise. We have $\left\langle \tilde{+}\tilde{-}\right.\left|\tilde{-}\tilde{+}\right\rangle =0$
and $\left\langle \tilde{\pm}\tilde{\mp}\right.\left|\tilde{\pm}\tilde{\mp}\right\rangle =1$.
The antisymmetric subspace is an invariant subspace to the collective
dephasing noise.

\section{Information masking}

Quantum information masking is a unique property of multi-body systems
in which quantum information is stored in the correlations between
subsystems rather than within them. It arises from the quantum entanglement
among these subsystems. The no-masking theorem proposed by Modi et
al. is one of the family of no-go theorems\cite{M}. Li et al. proposed
an interesting three-body information masking scheme that utilizes
the properties of Latin squares in mathematics\cite{Li}. Then we
simulated the quantum information masking process in Abelian anyon
system and Ising anyon system according to Li et al.'s Latin squares
scheme\cite{me2}. Here, using the mixed boundary punctures model,
Ising anyon statistics are simulated on toric code. We discuss the
Latin squares scheme of quantum information masking process under
this model.

In this case, we need three pairs of mixed boundary punctures (six
holes Fig.6) which corresponds to three Ising anyons. In our previous
work, we proved that three partite quantum information masking scheme
is mapping the arbitrary state $\alpha\left|1\right\rangle +\beta\left|\psi\right\rangle +\gamma\left|\sigma\right\rangle $
as
\begin{equation}
\begin{array}{c}
\left|\Psi\right\rangle =\frac{1}{\sqrt{3}}[\alpha\left|111\right\rangle +\alpha\left|\psi\psi\psi\right\rangle +\alpha\left|\sigma\sigma\sigma\right\rangle \\
+\beta\left|1\sigma\psi\right\rangle +\beta\left|\psi1\sigma\right\rangle +\beta\left|\sigma\psi1\right\rangle \\
+\gamma\left|1\psi\sigma\right\rangle +\gamma\left|\psi\sigma1\right\rangle +\gamma\left|\sigma1\psi\right\rangle ].
\end{array}
\end{equation}
In this case, we can get $Tr_{AB}\left(\left|\Psi\right\rangle \left\langle \Psi\right|\right)=Tr_{AC}\left(\left|\Psi\right\rangle \left\langle \Psi\right|\right)=Tr_{BC}\left(\left|\Psi\right\rangle \left\langle \Psi\right|\right)=I/3$,
the masking process is accomplished, the information is stored in
the quantum correlation rather than the subsystems. Please refer to
our work for the specific proof process. All terms in the braiding
operation satisfy the masking requirement obviously except for the
three Ising anyons term $\left|\sigma\sigma\sigma\right\rangle $.
The braiding operation gives each term an additional phase which can
cancel each other in the calculation of $\left|\Psi\right\rangle \left\langle \Psi\right|$.
According to the special statistical properties Eq.(\ref{eq:3}) and
Eq.(\ref{eq:4}), the case of term $\left|\sigma\sigma\sigma\right\rangle $
is a little complicated. In the mixed boundary punctures model, we
assume three pairs of mixed boundary punctures to realize three Ising
anyon state $\left|\sigma\sigma\sigma\right\rangle $ .

We give an example of the state $\left|\pm\pm\pm\right\rangle $,
other cases $\left|\pm\mp\pm\right\rangle $ and $\left|\pm\pm\mp\right\rangle $
are same.
\begin{equation}
\begin{array}{cc}
\left|\pm\pm\pm\right\rangle = & +\left|e_{1}e_{2}e_{3}e_{4}e_{5}e_{6}\right\rangle \pm\left|e_{1}e_{2}e_{3}e_{4}m_{5}m_{6}\right\rangle \\
 & \pm\left|e_{1}e_{2}m_{3}m_{4}e_{5}e_{6}\right\rangle +\left|e_{1}e_{2}m_{3}m_{4}m_{5}m_{6}\right\rangle \\
 & \pm\left|m_{1}m_{2}e_{3}e_{4}e_{5}e_{6}\right\rangle +\left|m_{1}m_{2}e_{3}e_{4}m_{5}m_{6}\right\rangle \\
 & +\left|m_{1}m_{2}m_{3}m_{4}e_{5}e_{6}\right\rangle \pm\left|m_{1}m_{2}m_{3}m_{4}m_{5}m_{6}\right\rangle .
\end{array}
\end{equation}
We assume $M_{1}=\left|e_{1}e_{2}e_{3}e_{4}e_{5}e_{6}\right\rangle \pm\left|m_{1}m_{2}m_{3}m_{4}m_{5}m_{6}\right\rangle $,
$M_{2}=\pm\left|e_{1}e_{2}e_{3}e_{4}m_{5}m_{6}\right\rangle +\left|m_{1}m_{2}m_{3}m_{4}e_{5}e_{6}\right\rangle $,
$M_{3}=+\left|e_{1}e_{2}m_{3}m_{4}m_{5}m_{6}\right\rangle \pm\left|m_{1}m_{2}e_{3}e_{4}e_{5}e_{6}\right\rangle $
and $M_{4}=\pm\left|e_{1}e_{2}m_{3}m_{4}e_{5}e_{6}\right\rangle +\left|m_{1}m_{2}e_{3}e_{4}m_{5}m_{6}\right\rangle $.
The braiding operation between Ising anyons $1$ and $3$ named $B_{13}$
gives $M_{3}$ and $M_{4}$ a minus sign and keeps $M_{1}$ and $M_{2}$
invariant. Similarly, the braiding operation between Ising anyons
$3$ and $5$ named $B_{35}$ gives $M_{2}$ and $M_{4}$ a minus sign
and keeps $M_{1}$ and $M_{3}$ invariant. $B_{13}$ and $B_{35}$
are both adjacent braiding of Ising anyons. It is not difficult to
prove, in the adjacent braiding case, the minus signs can offset each
other. There is another more complicated braiding type, three partite
braiding. For instance, braiding $\sigma_{1}$ around $\sigma_{2}$
and $\sigma_{3}$ corresponds to braiding $s_{1}$ around $s_{3}$
and $s_{5}$ . Three partite braiding which has the same beginning
and end particle positions has two styles (Fig.7). In Fig.7(a), particle
$2$ and $3$ do not braid, $B=B_{13}B_{15}B_{51}B_{31}$ is composed
of four adjacent braiding. In Fig.7(b), particle $2$ and $3$ also
braids, the diagram has two links. If we define the braiding $B^{'}$
is a cyclic permutation, then $B^{'3}$ gives the same extreme points
of the strands. These two braiding styles are quite different obviously,
which makes the problem more difficult. Fortunately, the three partite
system is Ising anyon system. In Yu's research\cite{Y}, it has been proven
that the braiding of Ising anyons is only related to the endpoints
of strands, since the braiding group is a subgroup of the Clifford
group, which is a finite group. It is noteworthy that this conclusion
applies only to Ising anyons. The braiding groups of other anyons
are infinite groups and heavily depend on the manner of braiding.
Thus the two diagrams in Fig.7 are the same $B=B^{'}$ here, we can conclude
them as Fig.8 in which the braiding process is a black box. Therefore,
the braiding of three partite is equivalent to the successive action
of adjacent braiding which can realize the quantum information masking
as mentioned above.
\begin{figure}
\includegraphics[scale=0.3]{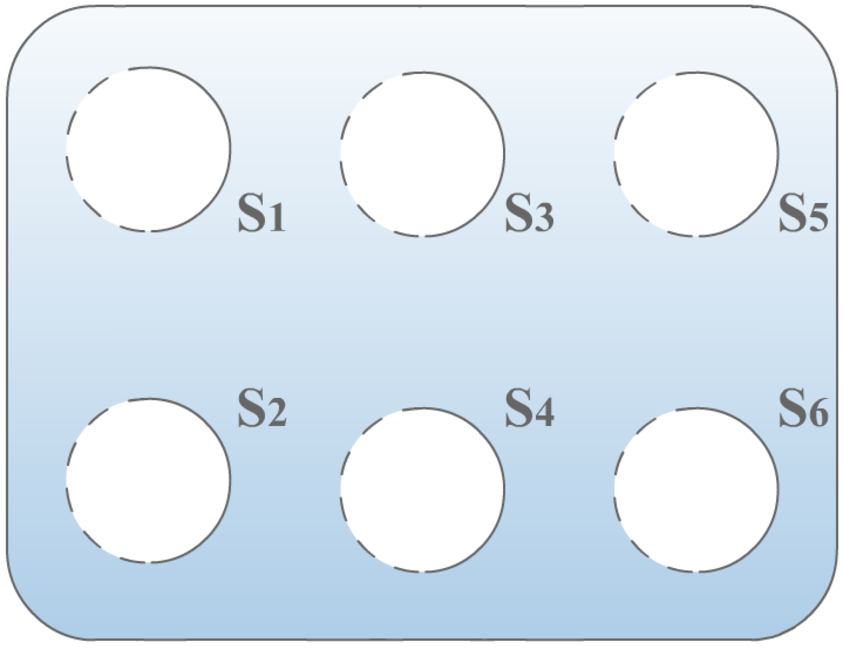}\caption{Six mixed boundary punctures model simulates three Ising anyon state.}
\end{figure}

\begin{figure}
\subfigure[Four adjacent braiding]{\includegraphics[scale=0.7]{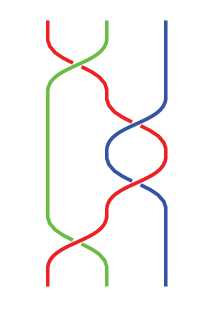}}\subfigure[Braiding includes two links.]{\includegraphics[scale=0.7]{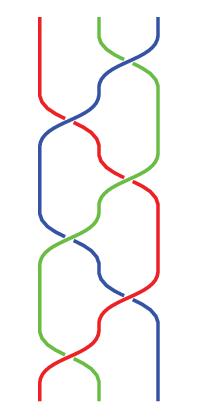}}\caption{Two different braiding styles give the same extreme points of the
strands. These two distinct braiding styles lead to two fundamentally different definitions of identity.}
\end{figure}

\begin{figure}
\includegraphics[scale=0.7]{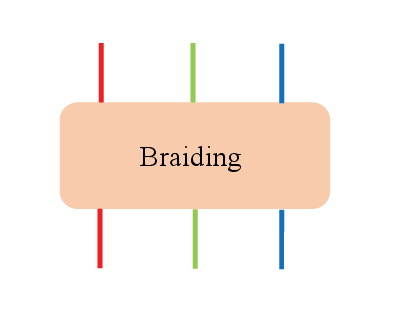}\caption{The simplified model of braiding in Ising anyon system. The braiding of Ising anyons is only related to the endpoints
of strands, since the braiding group is a subgroup of the Clifford
group \cite{Y}.}
\end{figure}

\section{Summary}

The toric code, also known as the surface code, is a renowned topological anyon model proposed by  Kitaev.
Defects within the toric code can exhibit non-Abelian statistical properties, enabling the potential for topological quantum computation.
The defects can be classified into punctures and twists.
Benhemou \emph{et al.} hybridized these two types of defects into the mixed boundary punctures model, combining their advantages.
 This mixed boundary punctures model demonstrates the non-Abelian statistics of Ising anyons.
 They proposed that non-Abelian statistical properties could be realized within the symmetric subspace,
 but we have demonstrated that the antisymmetric subspace can also exhibit these properties.

Given that this mixed boundary punctures model can simulate the statistical properties of Ising anyons on the toric code, we investigate whether it also possesses fault tolerance similar to that of non-Abelian anyons. We delve into two typical types of noise and find that the mixed boundary punctures model is immune to these disturbances. Furthermore, we demonstrate that this system can achieve quantum information masking by storing information in the correlations between subsystems rather than within the subsystems themselves.

\section*{Acknowledgement}

The research was supported by the Fundamental Research Funds for the
Central Universities, China No.2022JKF02024, the National Natural
Science Foundation of China No.11675119.

\end{document}